# Gravity Wave Phase Shift in a Cold Quark Star with a Nonconvex QCD BZT Shock Wave Van Der Waals Equation of State

**Keith Andrew [1,*], Eric V. Steinfelds [2] and Kristopher A. Andrew [3]**

[1] Department of Physics and Astronomy, Western Kentucky University Bowling Green, KY 42101, USA
[2] Department of Computational and Physical Sciences, Carroll University, Waukesha, WI 53186, USA; eric.steinfelds@wku.edu
[3] Department of Science, Schlarman Academy, Danville, IL 61832, USA; kandrew@schlarman.com
* Correspondence: keith.andrew@wku.edu

**Abstract**

We investigate BZT shocks and the QCD phase transition in the dense core of a cold quark star in beta equilibrium subject to the multicomponent van der Waals (MvdW) equation of state (EoS) as a model of internal structure. When this system is expressed in terms of multiple components, it can be used to explore the impact of a phase transition from a hadronic state to a quark plasma state with a complex clustering structure. The clustering can take the form of colored diquarks or triquarks and bound colorless meson, baryon, or hyperon states at the phase transition boundary. The resulting multicomponent EoS system is nonconvex, which can give rise to Bethe–Zel'dovich–Thompson (BZT) phase-changing shock waves. Using the BZT shock wave condition, we find constraints on the quark density and examine how this changes the tidal deformability of the compact core. These results are then combined with the TOV equations to find the resulting mass and radius relationship. These states are compared to recent astrophysical high-mass neutron star systems, which may provide evidence for a core that has undergone a quark gluon phase transition such as PSR 0943+10 or GW 190814.

## 1. Introduction

The study of the interacting dynamics of cold quark matter in beta equilibrium to understand a portion of the QCD phase diagram to describe the existence of compact stellar objects has grown considerably with the discovery of anomalous events such as the stellar remnant from the energetic hypernova SN 2006gy [1]. This event, caused by the sudden collapse of a high-mass star, gave rise to final state mass and radius values that can differ from standard compact stellar core models. The ongoing observational datasets now include the anomalous X-ray pulsar [2] RX J1856 [3,4] and pulsars such as PSR 0943+10 [5], PSR B1828-11 [6], PSR Jo751+1807 [7], PSR B1642-03, PSR J1614-2230 and PSR J0348-0432 along with potential, GRB candidates GRB 050709 [8] and GR B170817A [9,10] and magnetar candidates such as 1E 2259+ 586 [11,12], and gravity wave events [13,14] where masses, radii, dual luminosity peaks and cooling rates do not always give extraordinary fits to conventional models based upon degenerate neutron matter.

Improvements in the observational data have resulted in more detailed models to help understand the varied mechanisms at play in dense QCD matter beyond the nuclear saturation limit of ~0.16 $fm^{-3}$. New models [15] are gaining support from the observations of the Quark Gluon Plasma, QGP, demonstrating the existence of a high temperature, low chemical potential state of unconfined quarks as seen at SPS [16], LHC [17], and RHIC [18,19] laboratories. In recent years, the QGP data have expanded to include the determination of relativistic QCD-based transport coefficients [20] and measurements of the overall bulk viscosity for free quark matter [21,22]. Observations of X-ray, $\gamma$-ray, and quasi-periodic objects such as SAS J1808.4-3658 which has a 2.49 ms period [23] have been used to understand how the environment [22] and the ambient mass density impact fundamental QCD parameters and potentials [24]. Various EOS have been used linking the strong coupling constant and quark masses to the local density function using the real-time formalism and the Dyson–Schwinger gap equation at finite temperature [25]. In this way, high-density compact cores can be used to give the density-dependent strong coupling that can be expressed in a way that is suitable for EOS and phase transition analysis [26] with high chemical potentials and lower temperatures. The high-resolution data from the NICER satellite have extended and constrained the EoS modeling with X-ray spectra [27] for masses and radii [28].

For a cold collapsing star in beta equilibrium the overall charge, density, and chemical potentials are constrained by overall charge neutrality for particle number density $n_j$, mass density $\rho_f$ and electric charge $q_j$ for particle type $j$ or flavor $f$: $\sum q_f n_f = q_e n_e$ and $\rho_B = \rho_u + \rho_d + \rho_s$ and three light quark flavor weak interaction equilibrium from quark interactions $d \leftrightarrow u + e^- + \bar{\upsilon}_e$, $s \leftrightarrow u + e^- + \bar{\upsilon}_e$, $s + u \leftrightarrow d + u$ constraining the chemical potentials where the neutrinos and antineutrinos exit the core on a short time scale compared to the core formation process effectively causing their chemical potentials to vanish resulting in $\mu^d = \mu^u + \mu^e$, $\mu^d = \mu^s$. The classical pressure is related to the chemical potential from $P = -\partial U/\partial V = n^2 \partial(\varepsilon/n)/\partial n = n\mu - \varepsilon$ where the chemical potential is $\mu = \partial\varepsilon/\partial n$. For the massless quark case this gives a direct comparison to the MIT bag model [29] with bag constant B where $3P = (\varepsilon - 4B) = \varepsilon - 4(3\mu^4/4\pi)$ linking [30] the vacuum pressure bag constant to the quark chemical potentials which are energy dependent. Several compact core models have focused on modified EOS for various states of dense matter in the QCD ground state of Color Flavor Locked (CFL) dense matter [31,32] with 2-D and 3-D color superconductivity [33–35] and in the Nambu-Jona-Lasinio (NJL) model [36]. Many models make a comparison to versions of the generalized MIT Bag model [37] approach and include QCD mean field approximations [38,39] quarkyonic matter [40,41] along with computationally intensive numerical lattice approaches that are being improved to accommodate finite temperature chemical potentials [42]. Refined models include crustal effects [43], deformations [44], layering [45], cooling rates [46], strange quark cores, finite temperature boson stars [47] and kaon condensation [48,49] superconducting color vector potential effects [50,51], phase transitions, and different crystalline cores [52–54]. In many cases these EoS are everywhere convex, however, there are a number of cases where the pressure and density are so extreme that the EoS will become nonconvex. In the case of relativistic hydrodynamics [55,56] the energy momentum tensor [57] has a term proportional to the enthalpy which includes a factor of $P/\varrho c^2$, or a factor of $4B/\varrho c^2$ for the MIT Bag model, which is normally quite small. At high pressures this term makes a larger contribution and, when asymptotic freedom is considered leading to free fast-moving particles, this term can be significant. At first this correction will drive the system closer to a stable point, but for high core pressures it can continue to push the system into an unstable region of shock waves that will alter the phase transition. Such phase transitions, occurring in the nonconvex region of the EoS, have been studied by Bethe [58], Zel'dovich [59] and Thompson [60] and these fluids are now known as BZT fluids. They developed a fundamental derivative test to identify when a fluid could develop BZT behavior. This has been used in the study of neutron stars by Aloy et al. [61] where they develop a method for measuring the BZT signature in gravitational waveforms [62,63]. The BZT phase transitions can have regions of spinodal decomposition leading to the formation of quark clusters and diquark states from the color states for color group SU(3)C $3 \otimes 3 = \overline{3} + 6$, where the antisymmetric color factor is attractive, C(F) < 0, and the symmetric is a weaker repulsive force, C(F) > 0. Here we will apply the multicomponent van der Waals model to investigate the hadron to quark phase transition in a compact stellar core in beta equilibrium with a BZT region identified by the fundamental derivative test. As noted by Ghosh [64] and Vovchenko [65] where they investigate the nuclear models and van der Waals [66] correspondence from the resonance hadron gas mode to lattice QCD, this model provides an analytical model with a critical point, it can model composite quark clusters as color singlets or net color carriers [67], it includes a phase transition, it can support shock wave behavior, it includes an excluded volume limit, the parameters are selected in alignment with lattice models, and can include both a repulsive or attractive strong force limit using color factor couplings depending upon the color group representation of SU(3)C. As such several authors have used the van der Waals (vdW) equation [68–72] of state (EoS) provides a useful analytical framework to describe the complex interactions within a hadronic medium under extreme conditions [73–77].

However, the MvdW EoS phase transition zone admits a nonconvex region that allows for BZT shock waves to form. When the BZT fundamental derivative becomes negative, the conditions for nonclassical anomalous wave formation exist. The BZT compression can lead to a rapid dispersion and an undefined speed of sound. In this case, the nonconvex MvdW EoS [78,79] can be used to describe the BZT phase transition as being spinodal. Due to the nonclassical behavior of shock waves in this region, spinodal clumping of mixed phases is expected to occur [80]. Israel has recently developed a relativistic generalization of these BZT fundamental derivative results, opening a new regime to explore. As a result, the compression can be purely classical, or a nonclassical BZT-type, or a nonclassical relativistic BZT-type. This nonlinear wave propagation and dispersion leads to a shift or reduction or phase shift in gravity wave production whose signature can be searched for using the Laser Interferometer Gravitational Wave Observatory (LIGO), the VIRGO Gravity Wave Interferometer, the Kamioka Gravitational Wave Detector (KAGRA), the Einstein Telescope gravitational wave detector (ET) or the Laser Interferometer Space Antenna (LISA) data or simulations. In general, shock waves in a BZT hadron fluid with a negative fundamental derivative could contribute to a QCD phase transition by creating nonclassical localized high-density regions conducive to quark formation facilitated by quark nuggets or noncolor singlet quark clusters. This could result in a broader and more easily sustained phase transition to a stellar quark core, leaving a unique gravity wave or luminosity signature that can be partially modeled by a MvdW EoS. This paper is organized as follows: the next section introduces the multicomponent van der Waals partition function and the various thermodynamic functions used to calculate the fundamental derivative and its relativistic correction with a special focus on the BZT regions, the following section models the gravitational wave signature of the BZT zone through the tidal deformability modeled from solutions to the TOV equation with the MvdW EoS, finally these are compared to the sensitivity range needed for detection, finally ending with a concluding section. We use units where G = c = k = 1 and the lowercase letter for thermodynamic variables that are specific densities per volume.

## 2. MvdW Partition Function and Fundamental Derivative

Here, we calculate the thermodynamic quantities needed to derive the BZT fundamental derivative for the MvdW EoS. Starting with the partition function, we will calculate the internal energy and enthalpy for our system. The MvdW partition function [81,82] for a total of $N_c$ different components, where each component species can have $N_i$ particles, each with a van der Waals volume of $b_k$, with a mean field interaction coupling strength given by $a_{ij}$, in a total volume V, at temperature T, is given by:

$$Z_{vdw}\left(N_i,V,T\right)=\prod_{i=1}^{N_C}\frac{1}{N_i!}\left(\frac{V-\sum_{j=1}^{N_C}N_j b_j}{\lambda_i^3}\right)^{N_i}\exp\left(\frac{N_i}{VkT}\sum_{j=1}^{N_C}a_{ij}N_j\right) \tag{1}$$

where the i[th] particle species has a thermal de Broglie wavelength given by:

$$\lambda_i=\sqrt{\frac{1}{2\pi m_i T}} \tag{2}$$

For our applications, the interaction couplings will obey a mixing rule where $a_{ij}$ is symmetric and can be considered as the product of the individual particle couplings $a_i$ where we use the interaction parameter $k_{ij}$ expressed as:

$$a_{ij}=\sqrt{a_i a_j}\left(1-k_{ij}\right)=a_{ji} \tag{3}$$

In general, this expression plays the role of a mixing rule for different components where the dimensionless $k_{ij}$ gives a binary interaction parameter as in Keffer [82], and several chemical examples can be found in Stryjek [83]. Different authors have used this term to introduce color force asymmetry, flavor structure differences between quarks, or to model hadronic diquark and gluon interactions. The $k_{ij}$ term can be modeled as a (u, d, s) 3 × 3 matrix with diagonal terms reflecting stronger QCD interactions and with the off-diagonal terms of varying interaction strengths. In this context, the $a_{ij}$ can represent the QCD Cornell potential, giving the direct channel strength, and the $k_{ij}$ gives the indirect and nonsinglet mixing strengths for different diquark pairs. For example, in composite systems with diquarks or hexaquark states, the (u, d, s) system $k_{uu}$ = $k_{dd}$ = $k_{ud}$ = 0.2, while $k_{us}$ = $k_{ds}$ = 0.4 and in the CFL color (r-red, g-green, b-blue) state $k_{br}$ = 0.1 when $k_{rg}$ = 0.3 from gluon color suppression in states of (r, g, b) [84,85]. We see that as $k_{ij}$ increases, the effective interaction strength decreases, thereby softening the EOS. Due to this effect, we will set $k_{ij}$ = 0 for our calculations. The values of the $a_{ij}$ terms can reflect the different color factor interactions and can represent attractive or repulsive interactions. The pressure can be expressed in terms of particle species number $N_i$ or number volume density:

$$\begin{gathered}p=T\left(\frac{\partial \ln Z}{\partial V}\right)_{N,T}=T\sum_{i=1}^{N_c}\left[\frac{N_i}{V-\sum_{j=1}^{N_c}N_j b_j}-\frac{N_i}{V^2T}\sum_{j=1}^{N_c}N_j a_{ij}\right]\\ p=\frac{\sum_{i=1}^{N_C}n_i T}{\left(1-\sum_{j=1}^{N_C}n_j b_j\right)}-\sum_{i=1}^{N_C}\sum_{j=1}^{N_C}n_i n_j a_{ij}\ .\end{gathered} \tag{4}$$

The internal energy is given by:

$$U = kT^2\left(\frac{\partial \ln Z}{\partial T}\right)_{N,V} = T\sum_{j=1}^{N_C} N_j - \frac{1}{V}\sum_{i=1}^{N_C}\sum_{j=1}^{N_C} a_{ij}N_iN_j \tag{5}$$

and the entropy per particle is given by:

$$S_p = S_0 + \sum_{i=1}^{N_C}\frac{N_i}{N}\ln\left(\frac{V_p - b_{mix}}{\Lambda_i^3}\right) - \sum_{i=1}^{N_C}\frac{N_i}{N}\ln\left(\frac{N_i}{N}\right) \tag{6}$$

Which will be used to express the pressure and the classical enthalpy in terms of entropy and volume to evaluate along an entropic curve, and the enthalpy volume density h, is:

$$h = u + pV = \frac{3}{2}\frac{N}{V}T - 2\sum_{i=1}^{N_C}\sum_{j=1}^{N_C} a_{ij}\frac{N_i}{V}\frac{N_j}{V} + \frac{T\left(N/V\right)}{1 - \sum_{j=1}^{N_C}\frac{N_j}{V}b_j} \tag{7}$$

We simplify these expressions by introducing a scale factor based on the total number of particles, N, for each particle species number fraction $x_i$, the total van der Waals mixture volume of the particles, $b_{mix}$, and mixture interaction couplings, $a_{mix}$, for N particles with specific heat $c_V$, with reciprocal δ:

$$\begin{gathered} N = \sum_{k=1}^{N_C} N_k, \quad x_i = \frac{N_i}{N}, \quad V_p = \frac{V}{N}, \quad b_{mix} = \sum_{k=1}^{N_C} x_k b_k, \\ a_{mix} = \sum_{j=1}^{N_C}\sum_{k=1}^{N_C} x_j x_k a_{jk}, \quad n_i = \frac{N_i}{V}, \quad \delta = \frac{1}{c_V}, \end{gathered} \tag{8}$$

where the specific heat $c_V$ is related to the number of degrees of freedom for the ith particle species. Using these variables, the multicomponent van der Waals EoS and specific entropy and energy density ε become:

$$\begin{gathered} p(V,T) = \frac{RT}{V_p - b_{mix}} - \frac{a_{mix}}{V_p^2}, \quad \varepsilon = \frac{RT}{\delta} - \frac{a_{mix}}{V_p} + \varepsilon_0, \\ s = R\ln\left[\left(V - b_{mix}\right)\left(\varepsilon + \frac{a_{mix}}{V}\right)^{\frac{1}{\delta}}\right] + s_0 \end{gathered} \tag{9}$$

where $s_0$ and $\varepsilon_0$ are integration constants set to reference values, for the cold beta equilibrium quark star. Changing variables from temperature and volume to specific entropy and volume we have:

$$P\left(T,V\right) \rightarrow P(s,V) = \frac{\delta\exp\left[\frac{\delta\left(s - s_o\right)}{R}\right]}{\left(V - b_{mix}\right)^{\delta+1}} - \frac{a_{mix}}{V^2} \tag{10}$$

The first classical fundamental derivative $G_{Cl}$ is by definition:

$$G_{Cl} = -\frac{1}{2}V\frac{\left(\frac{\partial^2 P}{\partial V^2}\right)_s}{\left(\frac{\partial P}{\partial V}\right)_s} \tag{11}$$

where the multicomponent van der Waals classical BZT first fundamental derivative along an isentrope is given by:

$$G_{Cl} = \frac{(\delta+1)(\delta+2)V^5\left(P + a_{mix}/V^2\right) - 6a_{mix}V\left(V - b_{mix}\right)^2}{2V^4(\delta+1)\left(V - b_{mix}\right)\left(P + a_{mix}/V^2\right) - 4a_{mix}} \quad (12)$$

This gives the standard value in the limit of a single component and, when taking the ideal gas limit, we have a value of $G_{Cl} = 1 + \delta/2$ in agreement with the convex ideal gas law, in the limit of a single component we get the standard MvdW value, in the limit of no excluded volume we get the simplified ideal gas volume dependence and in the limit of no interactions with the excluded volume we get the interacting gas expression [86,87]. The locus of points (P, V) for G = 0 corresponding to the boundary between the classical and nonclassical regimes is found by solving for the pressure to obtain:

$$P(G=0) = \frac{a_{mix}}{V^2}\left[\frac{6}{(\delta+1)(\delta+2)}\left(1 - \frac{b_{mix}}{V_p}\right)^2 - 1\right] = n^2 a_{mix}\left[\frac{6\left(1 - nb_{mix}\right)^2}{(\delta+1)(\delta+2)} - 1\right] \quad (13)$$

The relativistic fundamental derivative is defined from the energy momentum tensor $T^{\mu\nu}$ and the specific enthalpy h. For fermions using the Fermi-Dirac distribution function f(p) for particles of mass m and Fermi momentum $p_F$ and the classical speed of sound squared $c_{Cl}^2$ defined along an isentrope for a system of total number density n, this is given by:

$$T^{\mu\nu} = nhu^\mu u^\nu + pg^{\mu\nu}, \quad c_{cl}^2 = \left(\frac{\partial P}{\partial n}\right)_S, \quad \text{where}$$

$$nh = \frac{g}{(2\pi)^3}\int\sqrt{p^2+m^2}\,f(p)d^3p - \sum_{i,j}a_{ij}n_in_j + \frac{\sum_i n_iT}{1-\sum_i b_in_i} - \sum_{i,j}a_{ij}n_in_j \quad \text{or} \quad (14)$$

$$nh = \frac{g}{8\pi^2}\left[p_F\sqrt{p_F^2+m^2}\left(2p_F^2+m^2\right) - m^4\ln\left(\frac{p_F+\sqrt{p_F^2+m^2}}{m}\right)\right] - \sum_{i,j}a_{ij}n_in_j + P.$$

For a cold star in beta equilibrium composed of nondegenerate fermions along an isentrope we use the classical speed of sound squared and the specific enthalpy to calculate the special relativistic contribution to the fundamental derivative $G_{sr}$, and add this to the classical fundamental derivative $G_{Cl}$ from Equation (12) to obtain the total fundamental derivative $G_{Total}$, this results in;

$$\begin{aligned}
nh &= 2\pi^{2/3}n^{4/3} - 2\sum_{i,j}a_{ij}n_in_j \\
G_{sr} &= -\frac{3}{2}c_{sr}^2 = -\frac{3}{2h}\left(\frac{\partial P}{\partial n}\right)_s \\
G_{Total} &= G_{cl} + G_{sr} = G_{Cl} - \frac{3c_{cl}^2}{2h} \\
G_{Total} &= G_{Cl} - \frac{6(\delta+1)\left(P+n^2a_{mix}\right) + 12n^2a_{mix}}{2\left(1-nb_{mix}\right)\left(\pi^{2/3}n^{4/3} - n^2a_{mix}\right)}.
\end{aligned} \quad (15)$$

Equation (15) can provide better treatment for the high-density behavior found in compact stellar cores, for equilibrated neutron stars well below the Fermi energy the temperature terms are very small and can be neglected for modeling, and the multicomponent nature of the MvdW EoS will allow us to explore the impact of having several particle species in the core where both attractive and repulsive interacting terms that can be tested using the interaction potential terms in the $a_{ij}$ and $k_{ij}$ factors. The relativistic term to the fundamental derivative value can shift a region that was an otherwise convex classical region into a new nonconvex BZT spinodal decomposition mixed fluid zone, thereby altering the phase change dynamics of the core. The clumping that develops from this phase change leads to increased nonclassical shocks and rarefactions whose signature can be apparent in the shift in the phase peak of the gravity wave spectrum. For positive classical fundamental derivatives, the relativistic correction can lower the value of the fundamental derivative closer to zero, thereby smoothing out fluctuations that could give rise to instabilities, as long as it does not become negative. In Figure 1, we plot several special cases of the BZT fundamental derivative and the relativistic correction for the multicomponent van der Waals EoS, where the negative regions correspond to the rarefaction shock wave zones.

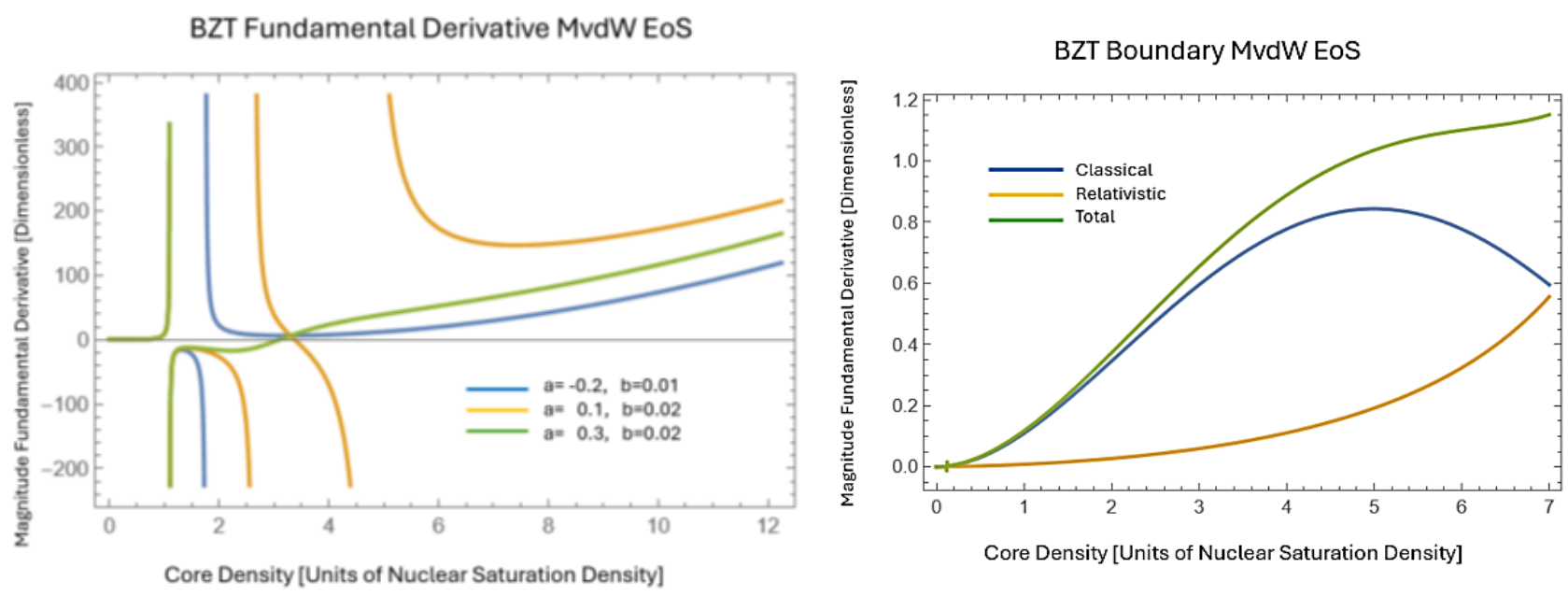

(**a**) (**b**)

**Figure 1.** (**a**) The first fundamental derivative for the multicomponent van der Waals equation of state showing positive and negative BZT regions for different strengths, densities, and relativistic corrections with no mixing, $k_{ij} = 0$, where going from blue to yellow to green is for increasing values of $a_{ij}$ and $b_k$. (**b**) The fundamental derivative boundary between the BZT zone and classical zone and the relativistic contribution to the total for $a_{mix} = 0.2$ GeV$^{-2}$ fm$^3$, $b_{mix} = 0.02$ fm$^3$, $a_{mix} = 0.1$ GeV$^{-2}$ fm$^3$, $b_{mix} = 0.02$ fm$^3$, and $a_{mix} = 0.3$ GeV$^{-2}$ fm$^3$, and $b_{mix} = 0.02$ fm$^3$ with $k_{ij} = 0$, where the blue curve is the classical contribution, the yellow curve is only the relativistic contribution and the green curve is the total.

To numerically evaluate Equations (12)–(15), we select the constituents to be point particles with three degrees of freedom, there are no degeneracies between states, and all densities are in units of nuclear saturation density. Nuclear saturation density, typically denoted $\varrho_0 \approx 2.7 \times 10^{14}$ g/cm$^3$ for mass density or $n_0 \approx 0.16$ fm$^{-3}$ for number density, represents the equilibrium density of symmetric nuclear matter inside atomic nuclei, where the attractive nuclear forces and short-range repulsion (including Pauli exclusion effects) are balanced. It serves as a fundamental reference point in compact star modeling, particularly for distinguishing the outer hadronic layers from the ultra-dense quark core in hybrid or pure quark stars [88]. In quark star density modeling, saturation density marks the threshold above which nucleonic degrees of freedom may dissolve into deconfined quark matter, depending on the stiffness of the equation of state and the nature of the phase transition (e.g., Maxwell, Gibbs, or BZT-type). Most quark star models—especially those incorporating a mixed phase—are calibrated to begin deviations from purely hadronic behavior at or just above $n_0$, making it a key boundary for initiating BZT-type transitions with rarefaction shocks.

The values for the MvdW parameters of $a_{mix}$ and $b_{mix}$ are heavily constrained from quark modeling of hadrons. The attractive interaction strength $a_{mix}$ reflects the effective color-mediated coupling between different quark flavors and is inspired by $SU(3)_C$ Casimir scaling for gluon exchange in the one-gluon exchange approximation. This allows us to encode the fact that interactions among red, green, and blue color charges are not identical but can be averaged over pairwise channels with weights tied to the antisymmetric color-singlet or symmetric color-octet configurations. In the present model, we start with the hadronic values $0.2 < a_{mix} < 0.6$ GeV$^{-2}$ fm$^3$, consistent with the effective coupling strengths used in Vovchenko et al. [65,71], who developed a van der Waals extension for Hadron Resonance Gas (HRG) and quark hadron crossover models. The extension of the interaction strength to the high core densities here is realized through the running coupling constant, which, for QCD, reduces the strong interaction force via asymptotic freedom. The repulsive excluded volume term $b_{mix}$ fm$^3$ represents the effective finite size of quark composites or effective repulsion from Pauli blocking and short-range gluon interactions. In line with recent work by Ghosh [64,89], who explored van der Waals quark matter in hybrid EoS construction, we take $0.7 < b_{mix} < 2.5$ fm$^3$, corresponding to an effective hard-core quark radius of 0.3–0.8 fm. In this sense,l $b_{mix}$ encodes repulsive interactions similar to a scattering impact parameter and is related to the radius of the underlying baryonic particle by $3\ b_{mix} = 16\pi r^3$. These choices are not arbitrary but lie within the parameter space explored in lattice-constrained hadron and quark matter simulations, providing a realistic interpolation between nuclear saturation and deconfined quark phases. Restrictions also come from the high-density fermion chemical potentials which have a more natural setting in the MIT bag model and the NJL extended model. The case of net color diquarks and larger quark clusters can extend the range of $a_{mix}$ and $b_{mix}$ beyond the standard baryonic domain.

## 3. Peak Gravity Wave Shift Signature of BZT Shocks and Chirps

The spinodal decomposition during the phase change results in anomalous dispersion and shock wave formation that will alter the gravity wave signature of a wave-producing event with a BZT compact stellar core. The gravitational wave amplitude varies as $h(f) \sim (M_{ch})^{5/3} f^{-7/6}/D$, for chirp mass $M_{ch}$ defined in Equation (18), frequency and distance D, these changes are currently beyond detectability, however the phase shift for a BZT EoS is closer to the range of detectability so we will analyze it more closely. We determine the mass and radius values from the TOV equations [90–92] applied to a static spherically symmetric mass:

$$\frac{dP}{dr} = \frac{Gm\rho\left(1+\frac{P}{\rho}\right)\left(1+4\pi r^3\frac{P}{m}\right)}{r^2\left(1-\frac{2m}{r}\right)}$$
$$\frac{dm}{dr} = 4\pi r^2\rho \qquad (16)$$

which we solve numerically for the MvdW EoS for different values of the parameters $a_{mix}$, $b_{mix}$, $k_{ij}$, and $N_c$ with plots presented in Figure 2. To solve the Tolman–Oppenheimer–Volkoff (TOV) equations for relativistic stellar structure, the Mathematica function NDSolve is employed with an adaptive numerical integration scheme [93,94]. This method transforms the coupled differential equations for mass and pressure—governed by the energy density profile and the equation of state (EOS)—into a numerically tractable system. The integration begins at a small initial radius with a specified central pressure and proceeds outward until the pressure drops to zero, marking the star's surface. NDSolve automatically selects a suitable integration method, typically using an embedded Runge–Kutta scheme for non-stiff regions and switching to backward differentiation formulas (BDF) in stiff regimes, as needed. The solver's adaptive step-size control ensures that local truncation errors remain within tolerance, adjusting the step size dynamically based on the complexity and steepness of the solution. This approach is essential for capturing the nonlinear behavior of the TOV system, particularly in the presence of realistic equations of state that may include phase transitions or nonconvex features such as BZT regions. The result is a smooth and accurate profile of the star's internal structure, including total mass and radius, across a wide range of central densities and EOS parameters.

The adiabatic index Γ, serves as a critical diagnostic tool in modeling the equation of state (EoS) for compact stars, bridging microphysical interactions with macroscopic stability and observational signatures. In all the above studies, Γ is employed to probe the stiffness of the EoS and the response of matter under compression, with its value determining the onset of dynamical stability, phase transitions, and wave propagation behavior. Moustakidis [95] uses Γ to define a critical threshold for the stability of relativistic stars, linking it analytically to compactness and central pressure via Chandrasekhar's variational method. Casali and Menezes [96] show how Γ decreases with the emergence of exotic particles like hyperons or strange quarks, indicating EoS softening and potential instability. Carney et al. [97], while not analyzing Γ explicitly, rely on its underlying structure in reconstructing the EoS from gravitational wave data, showing that smooth variations in Γ are essential to accurate spectral fits. Here we take a more dynamic view, showing how a drop in Γ—due to strong interparticle attraction, via $a_{mix}$, and limited excluded volume, via $b_{mix}$, in the multicomponent van der Waals EoS—can trigger nonconvex behavior, Bethe–Zel'dovich–Thompson (BZT) shocks, and gravitational wave phase shifts. Across these works, the adiabatic index functions as both a stability threshold and a marker for identifying nonlinear phenomena, making it indispensable for connecting theoretical models to observable astrophysical signatures. From the MvdW EOS, the adiabatic index is given by:

$$\Gamma = \frac{\partial \ln P}{\partial \ln \rho} > \frac{4}{3}$$
$$\frac{\partial \ln P}{\partial \ln \rho} = \frac{\left(1 - N_c n b_{mix}\right)\left[N_c T\left(1 + N_c n b_{mix}\right) - 2N_c^2 a_{mix} n \left(1 - N_c n b_{mix}\right)^2\right]}{\left(N_c T - a_{mix} N_c^2 \left(1 - N_c n b_{mix}\right)\right)} > \frac{4}{3} \tag{17}$$

Equation (17) is used to constrain the values of $a_{mix}$, $b_{mix}$, and n to be in a range that leads to a stable configuration. Expressed in terms of the number of components $N_c$, the particle number density n, the interaction strength $a_{mix}$, the excluded volume $b_{mix}$ which behaves like a repulsion under extreme compression, and the temperature T. The BZT shock zone from Equation (17) is plotted in Figure 2. The system becomes more stable as $a_{mix}$ is decreased and $b_{mix}$ is increased. Increasing $a_{mix}$ for higher densities mimics the asymptotic freedom effects seen in short-range QCD. Models with $a_{mix} > 0$ and $b_{mix} > 0$ exhibit monotonic increases as the density increases. However, the BZT fundamental derivative $G_{Total}$ is the more general diagnostic: it includes the adiabatic index and its variation and is required to determine whether the fluid dynamics will exhibit nonclassical features like rarefaction shocks or spinodal clumping, particularly relevant near phase transitions or in nonconvex EOS such as those arising in dense QCD matter. Meanwhile, Γ remains essential for basic stellar stability analysis but is not sufficient for capturing the full richness of nonlinear or anomalous wave behavior.

In general, the symmetry of the static and spherically symmetric stellar core would be broken by introducing rotation and magnetic fields which would be present in a collapsed stellar core. Including a magnetic field and the spin of the star alters the standard Tolman–Oppenheimer–Volkoff (TOV) solutions, which assume spherical symmetry and hydrostatic equilibrium in a non-rotating, unmagnetized compact object. Rotation breaks spherical symmetry and introduces centrifugal forces that counteract gravity, leading to a flattening of the star at the poles and an increase in equatorial radius. This deformation modifies the pressure gradient and mass distribution, requiring a generalization of the TOV equations to axisymmetric solutions, such as those described by the Hartle–Thorne approximation [98,99] or fully relativistic rotating star codes (e.g., RNS, https://github.com/cgca/rns, accessed on 5/2/2023). Magnetic fields, particularly those above $10^{15}$ G as found in magnetars or post-merger remnants, introduce anisotropic pressures due to the Lorentz force and the field's stress-energy tensor. The pressure becomes direction-dependent, with parallel and perpendicular components relative to the field lines, breaking isotropy and modifying the equilibrium structure. Moreover, strong magnetic fields can stiffen or soften the equation of state depending on orientation and particle magnetic moments, and they couple to the stellar currents and composition via Landau quantization and anomalous magnetic moment (AMM) effects [100,101]. Together, rotation and magnetic fields can increase the maximum mass and radius of the star compared to static TOV predictions, influence the star's moment of inertia, and affect gravitational wave signatures from oscillations or mergers. Here we first solve the static spherically symmetric case and look to add the impact of these terms in the future.

The Tolman–Oppenheimer–Volkoff (TOV) equations describe the hydrostatic equilibrium of relativistic stellar structures and yield distinct mass–radius solutions depending on the underlying equation of state (EOS). When the EOS is modeled using the multi-component van der Waals (MvdW) form, parameterized by the interaction strength $a_{ij}$ and excluded volume bk, each combination of $a_{ij}$ and $b_k$ produces a unique mass–radius curve with a corresponding maximum mass $M_{max}$ and radius $R_{max}$. These solutions can be characterized by the dimensionless compactness factor C = M/R, which quantifies the relativistic strength of the star's gravitational field. As the central density increases beyond the point where $M_{max}$ occurs, the solutions become unstable to radial perturbations, marking the onset of the classical TOV instability [102]—a purely gravitational instability tied to the loss of equilibrium. This behavior is fundamentally different from the BZT (Bethe–Zel'dovich–Thompson) shock-induced instability, which arises from nonconvex regions of the EOS where the fundamental derivative becomes negative, allowing anomalous shock propagation and spinodal decomposition. By systematically varying $a_{ij}$ and $b_j$, one can map out both the gravitational TOV instability boundary and the separate BZT instability zone, allowing a deeper understanding of how microphysical interactions in the EOS govern both equilibrium structure and dynamical behavior in compact stars, as shown in compactness and observational survey plot in Figure 2.

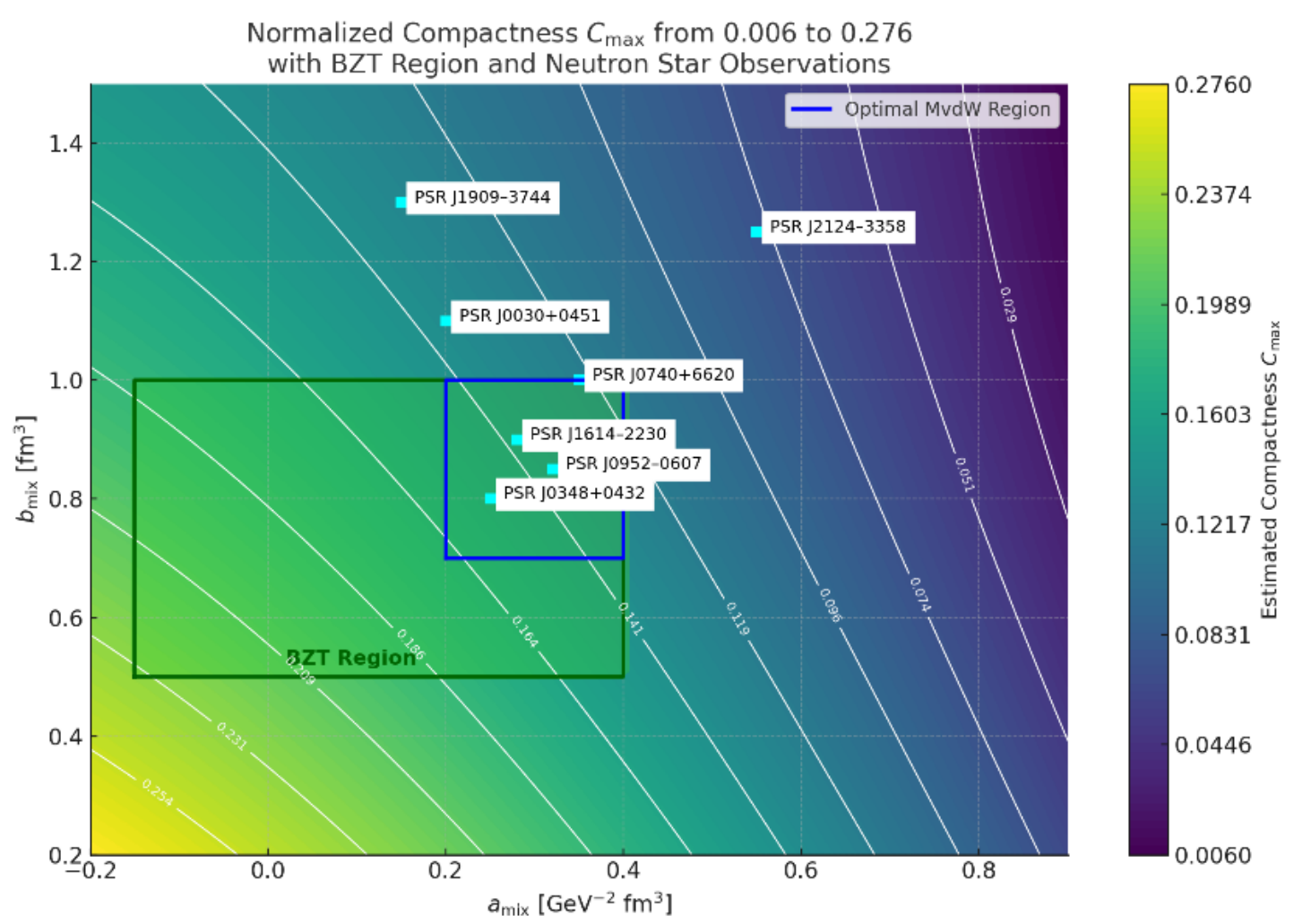

**Figure 2.** Maximum compactness $C_{max}$ = M/R over the multicomponent van der Waals (MvdW) EOS parameter space, shown as a color map (capped at 0.8) as a function of: $a_{mix}$ [$GeV^{-2}$ $fm^3$]—the strength of attractive interactions, $b_{mix}$ [$fm^3$]—the repulsive excluded volume parameter. White contour lines represent constant compactness from $0.04 < C < 0.28$, highlighting how EOS stiffness evolves across microphysical interaction space. Compactness is a critical structural parameter governing tidal deformability, gravitational wave phase shifts, and stellar stability. Higher compactness correlates with more compact and tightly bound neutron stars, while lower values are indicative of more deformable configurations. The larger rectangular boxed region denotes the BZT zone, defined by $-0.15 \le a_{mix} \le 0.4$ and $0.5 \le b_{mix} \le 1.00$, where the EOS exhibits thermodynamic nonconvexity (i.e., negative fundamental derivative $G < 0$), allowing the formation of compound shock structures and potentially richer dynamics in dense matter. The blue rectangle indicates the optimal region for the MvdW EOS based on physical viability, quantum-statistical consistency, and agreement with observations: $0.2 \le a_{mix} \le 0.45$ and $0.7 \le b_{mix} \le 1.0$ for the total range of values in the MvdW EOS.This range is compatible with quantum field theory-inspired models and provides causal, stable stellar solutions. Cyan squares mark the projected EOS parameter locations of several observed neutron stars: PSR J0030+0451, PSR J0740+6620, PSR J1614–2230, PSR J0952–0607, PSR J0348+0432, PSR J2124–3358, PSR J1909–3744, as referenced in Appendix C. Most stars lie within or near the optimal MvdW region, indicating these EOSs are consistent with both microscopic physics and macroscopic observables. Stars outside this region (e.g., J2124–3358, J1909–3744, as referenced in Appendix C) may probe EOS behaviors beyond QCD-calibrated interactions, offering valuable insight into the extremes of dense matter physics. For references on sources, see Appendix C.

The key implications for gravitational waves due to nonconvex QCD BZT shock waves during phase transitions lie in their ability to produce observable phase shifts in gravitational waveforms from binary mergers or neutron star oscillations. Nonconvexity in the equation of state (EoS), quantified by a negative fundamental derivative $G_{Total} < 0$, leads to anomalous wave dynamics, including compound rarefaction-compression shock structures. These features modify the internal density and pressure profiles of the star, altering the tidal deformability and gravitational wave phase evolution—effects that are potentially measurable by detectors such as LIGO, KAGRA, or LISA.

Figures 1–3 were generated using a multicomponent van der Waals (MvdW) EoS with parameters $0.4 < a_{mix} < 0.7$ $GeV^{-2}$ $fm^3$ and $0.5 < b_{mix} < 2.0$ $fm^3$, ranges motivated by lattice QCD analogs and color-mediated quark interactions. These values are realistic within the context of high-density QCD matter, where short-range repulsion (encoded in $b_{mix}$) and long-range attraction (through $a_{mix}$) emerge from gluon exchange and color confinement physics. The phase shifts shown (on the order of $\Delta\phi \sim 0.002$–$0.03$ radians) align with the detection thresholds of current and next-generation gravitational wave observatories. These calculations use a semi-analytic shock model and EOS-driven TOV integration to simulate both gravitational and gamma-ray burst signals, allowing a connection between microphysical nonconvexity and observable astrophysical signatures.

In Figure 2, we show numerical solutions exhibiting the mass and radius limits. We also show two candidate pulsars: J1614-223 and J1909-3744 (Appendix C reference) which have estimated masses near 1.5 and 2.0 solar masses. The waveform shift is influenced by the tidal deformability, $\Lambda$, the compactness C, and the second Love number $k_2$, where compactness values are determined from the TOV equations for the MvdW EoS for different particle content, and for a merging binary system the system compactness scales with the chirp mass, the merger tidal deformability depends on the individual masses and tidal deformability and the gravity wave phase shift can be expressed in terms of the chirp mass and merger deformability as reviewed in Baiotti [103] and presented in Vines in post Newtonian form [104] as

$$C_j = \frac{M_j}{R_j}, \quad \Lambda_j = \frac{2}{3} k_2 C_j^{-5}, \quad M_{ch} = \frac{\left(M_1 M_2\right)^{3/5}}{\left(M_1 + M_2\right)^{1/5}},$$
$$\tilde{\Lambda}_{merger} = \frac{16}{13} \frac{\left(M_1 + 12M_2\right) M_1^4 \Lambda_1 + \left(M_2 + 12M_1\right) M_2^4 \Lambda_2}{\left(M_1 + M_2\right)^5}$$
$$k_{2_{MvdW}} = \frac{8}{5}\left[\frac{C_j^5 \left(1 - 2C_j\right)^2}{1 + a_{mix} C_j + b_{mix} C_j^2}\right],$$
$$\text{Phase-Shift} : \Delta\Psi = \frac{3}{128}\left(\pi M_{ch} f\right)^{-5/3} \tilde{\Lambda}_{merger} \tag{18}$$

where the phase shift factor $\Delta\Psi$, we are using is the largest and first term of the series as presented in Vines.

The compactness and tidal deformability depend on the TOV masses and radii found for each EoS, allowing for a comparison between the various models, where several cases are presented in Figure 3. The Love number decreases with density, and the tidal deformability decreases with compactness. For the MvdW EoS, the stronger interactions for a > 0 and smaller number of components give a higher compactness and stiffness, resulting in less deformability, while a lower excluded volume, b << 1, leads to a greater tidal deformability.

Typically, the quark star has a higher compactness due to the increased density, leading to a lower tidal deformability when compared to a neutron star prior to the phase transition. The range of values is presented below in Table 1.

**Table 1.** Here are the comparison values for a neutron star and quark star for compactness, Love number, tidal deformability, chirp mass, and the merger tidal deformability.

| **Parameter** | **Neutron Stars** | **Quark Stars** |
|---|---|---|
| Compactness ($C$) | 0.15-0.25 | 0.2-0.35 |
| Love Number ($k_2$) | 0.05-0.15 | 0.01-0.05 |
| Tidal Deformability ($\Lambda$) | 200-1000 | 10-100 |
| Chirp Mass ($\mathcal{M}$ in solar masses) | 1.18-1.20 | 0.96-1.15 |
| Merger Tidal Deformability ($\tilde{\Lambda}$) | 70-720 | 10-100 |

The wave phase shift depends upon the EoS and the chirp mass; for the BZT region, we plot this change in Figure 3. Overall, the stiffer EoS gives a smaller shift due to reduced tidal effects with high compactness which is most pronounced at low frequencies. In the MvdW EoS, this corresponds to the fewest components with the largest attractive force, a > 0, and smallest excluded volumes, b << 1. Likewise, higher chirp masses lead to greater compactness and smaller phase shifts. Each model EoS gives a phase shift in accordance with the associated stiffness of the model.

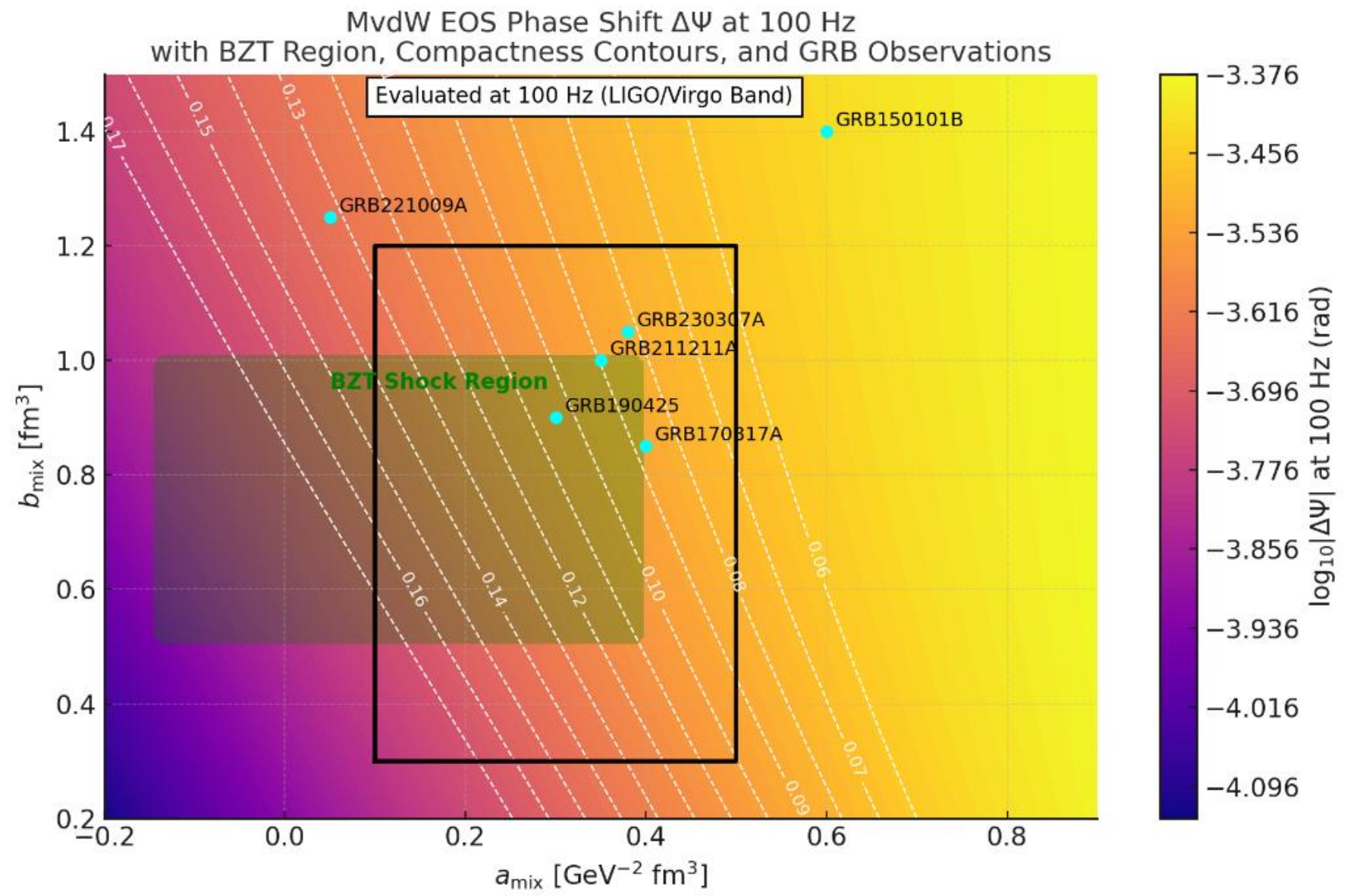

**Figure 3.** Gravitational wave phase shift $\Delta\Psi$ 100 Hz, shown as $\log_{10}|\Delta\Psi|$ (in radians), over the multicomponent van der Waals (MvdW) EOS parameter space. The plot spans the microphysical interaction parameters: $a_{mix}$ [GeV$^{-2}$ fm$^3$]: attractive interaction strength $b_{mix}$ [fm$^3$]: repulsive excluded volume. White dashed contours indicate lines of constant compactness $C = M/R$, a key structural property controlling tidal deformability and gravitational wave signatures. The color map encodes the strength of the phase shift $\Delta\Psi$, which reflects the EOS response to tidal fields in inspiraling neutron star binaries. The black rectangle bounds the QCD-motivated EOS region: $0.1 \le a_{mix} \le 0.50$, $0.3 \le b_{mix} \le 1.2$. These limits are based on lattice QCD and effective field theory models. The shaded region marks the BZT zone, where the EOS becomes thermodynamically nonconvex (negative fundamental derivative $G < 0$), enabling compound shocks and nonlinear wave behavior. This zone overlaps with many physically viable EOSs, and its proximity to observed GRB estimates suggests it may be astrophysically relevant. Cyan markers identify EOS locations associated with observed GRB-producing neutron star mergers, including: GRB170817A, GRB190425, GRB211211A, GRB230307A, GRB221009A, GRB150101B, see Appendix C for references. These events fall near or within the BZT region, within the phase shift detection threshold of LIGO/Virgo at 100 Hz—where their interferometric sensitivity is highest. Accurate modeling at this frequency is essential for resolving tidal effects and inferring EOS parameters from the gravitational waveform. While LIGO/Virgo is sensitive to phase shifts down to $\sim 10^{-2}$ rad, future detectors like LISA and the Einstein Telescope will push detection to $\sim 10^{-5}$ to $10^{-4}$ rad, requiring even finer constraints on EOS structure, especially in the low-compactness regime. This figure illustrates how GW phase observations, when mapped to microphysical EOS parameters, provide a bridge between QCD-scale interactions and astrophysical neutron star data. For references on sources, see Appendix C.

Figure 3 serves to demonstrate that the largest phase shifts arise in low-compactness, high-deformability regions and that several GRB-associated neutron star models (e.g., GRB170817A, GRB221009A, GRB150101B, references are in Appendix C) occupy regions near or within the BZT zone. The inferred values of $a_{mix}$ and $b_{mix}$ for these events are determined via matching gravitational wave phase shift, compactness, and tidal deformability observations to EOS model predictions. The alignment of observationally favored models with the BZT region underscores its potential relevance to real astrophysical systems.

The BZT regions reduce the energy going into the compression regions, thereby altering the coexistence energy distribution with anomalous dispersion and resulting in the peak energy to be delayed and frequencies to be shifted to lower values. A coherent picture emerges when Figures 2, 3, and A2 in Appendix A are considered together. Figure 2 demonstrates that the MvdW EoS supports compactness values consistent with known neutron star observations, such as PSR J0740+6620 and J0030+0451, thereby validating the model's structural realism. Figure 3 builds on this by showing that within the same parameter space, especially in the nonconvex BZT region, the MvdW EoS predicts gravitational wave phase shifts in the range detectable by current and next-generation observatories. These shifts are tied to tidal deformability and are especially relevant for GRB-associated neutron star mergers. Appendix A and Figure A2 reinforces the significance of the MvdW framework by comparing it to other EoS models, illustrating that it not only provides the strongest and broadest BZT shock wave structure, but also uniquely sustains a sizable energy discontinuity across the phase boundary—critical for modeling the internal dynamics and luminosity evolution of post-merger or hybrid PSR systems. Together, these figures establish the MvdW EoS as a reasonable candidate for describing some of the observable consequences of QCD phase transitions in compact stars.

For comparison, in Figure 4, we use modified BZT versions of the following models: the MIT bag model, the Hadron Gas model, the NJL model, and a lattice model. To introduce BZT regions we include interaction terms: in the MIT model we use the standard extension of a van der Waals like interaction with a bag constant [105,106], in the Hadron Gas Model the interaction term includes a field that repels and attracts [107], the NJL model has a BZT region for the chiral symmetry breaking that is a part of the model [108,109], the lattice model uses a perturbative QCD term that drives the pressure derivatives negative [110,111]. While it is possible to produce BZT regions in each model, the impact varies in extent and significance. In the modified MIT bag model, the BZT region exhibits a strong dependence on the magnitude of the interaction parameter, in the Hadron Gas model the BZT region only appears near hadronic saturation density and is highly localized, in the NJL model BZT effects are prominent and impactful in the chiral transition region, in lattice models BZT regions only occur near the critical point at high chemical potentials. The overall trend is similar for each BZT region, with a merging of all values for high chirp mass. We find that the MvdW EOS results in the largest phase shift $\Delta\Psi$, while the Lattice model produces the smallest shift, see Appendix A for details. A Bayesian Inference Corner Plot for the MvdW EOS parameters with posteriors is given in Appendix B, showing the parameter correlations.

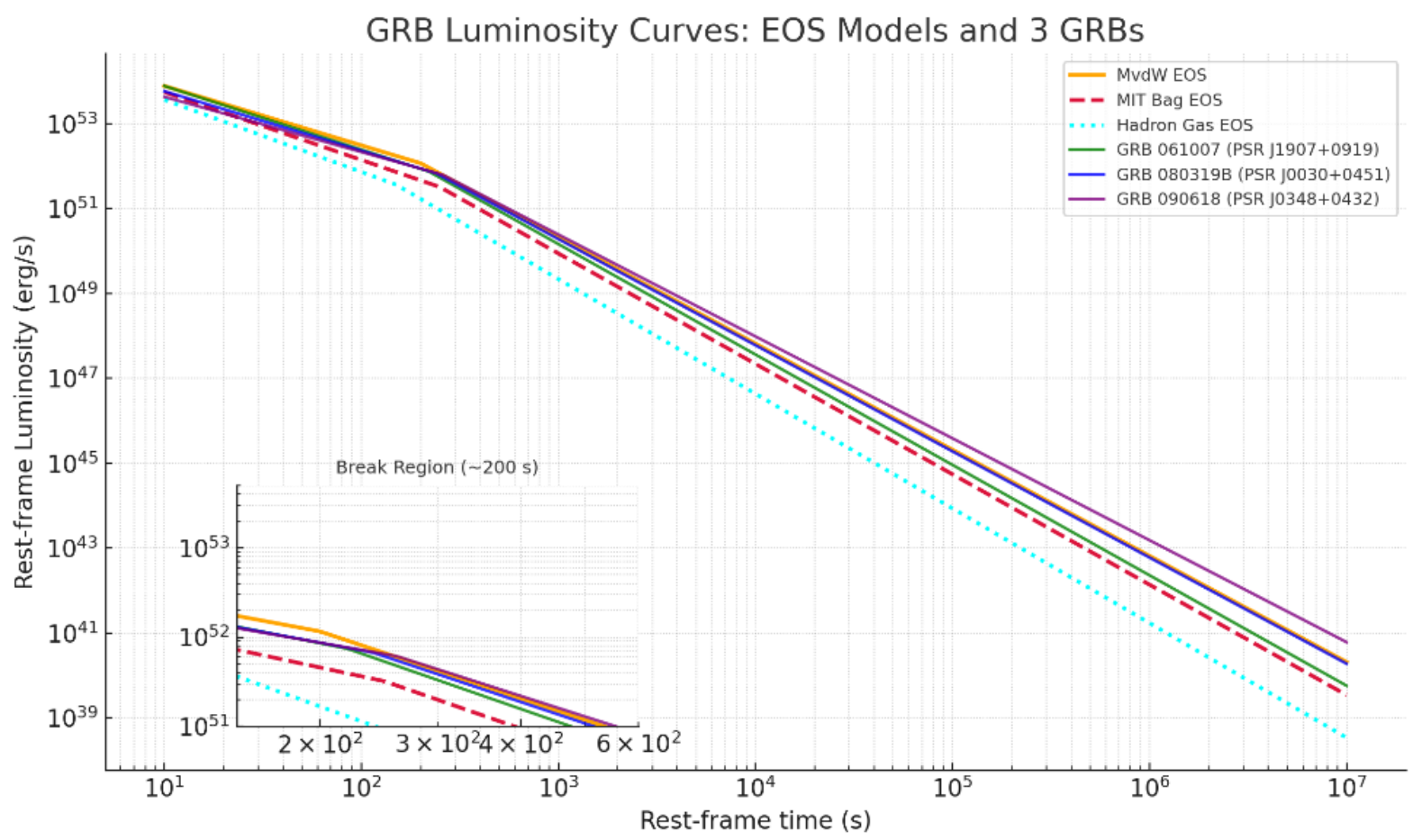

**Figure 4.** Comparison of GRB luminosity curves from three theoretical EOS models—Multicomponent van der Waals (MvdW), MIT Bag, and Hadron Gas—against observational light curves of three well-studied GRBs: GRB 061007, GRB 080319B, and GRB 090618 [139]. The curves are plotted as rest-frame luminosity versus time using broken power law fits. The MvdW EOS (orange) produces a broad, high-luminosity profile consistent with GRB 090618 (purple), while the MIT Bag EOS (red

dashed) matches the steep decay behavior of GRB 061007 (green). The Hadron Gas EOS (cyan dotted) yields a lower-luminosity evolution consistent with more rapidly fading bursts. The inset zooms into the critical break region near ~200 s, where GRB light curves transition from plateau to decay. These comparisons highlight the EOS sensitivity of BZT-driven GRB emissions and demonstrate how phase-transition energetics and shock dynamics map to GRB light curve morphology.

In Figure 5, we compare the values from Table 2 [112] with the phase shifts expected in Figure 4 with detector sensitivities to identify the thresholds for detection. The LIGO [113]-VIRGO [114]-KAGRA [115] detectors have phase shift sensitivities ~0.02 radians which would require MvdW parameters of $a > 4$ MeV fm$^3$ and $b < 0.01$ fm$^3$. These values do not result in a likely scenario at the required high densities, given that QCD forces exhibit asymptotic freedom at short distances. The upgraded Einstein Telescope [116] requires a~1.2 MeV fm$^3$ and b~0.04 fm$^3$ to produce observable phase shifts; these values are in the range expected during a BZT phase transition and could lead to a measurable result. The future LISA [117] detector should be in this range and could constrain the values used for the MvdW EoS.

**Table 2.** Here we collect the typical response characteristics of detectors that can find evidence for the predicted phase shift. The KAGRA data is for the current upgrade. Frequency in Hertz (Hz), Strain is the minimum detectable dimensionless strain amplitude as a measure of spacetime deformation, estimated signal to noise ratio (SNR) for each detector, and phase shift sensitivity is the minimum detectable phase distortion in radians.

| | Detector | Frequency Range (Hz) | Sensitivity (Strain) | SNR Threshold | Application | Phase Shift Sensitivity |
|---|---|---|---|---|---|---|
| 1 | LIGO | 10 –1000 | $\sim 10^{-23}$ | ~8 | Binary mergers, neutron stars, black holes | $10^{-2}$ |
| 2 | Virgo | 10 – 1000 | $\sim 10^{-23}$ | ~8 | Binary mergers, neutron stars, black holes | $10^{-2}$ |
| 3 | LISA | 0.01 – 1 | $\sim 10^{-21}$ | ~1 | Supermassive black holes, early universe | $10^{-5}$ |
| 4 | Einstein Telescope | 1 – 1000 | $\sim 10^{-24}$ | ~5 | Binary mergers, cosmological signals | $10^{-4}$ |
| 5 | KAGRA | 10 – 1000 | $\sim 10^{-23}$ | ~8 | Binary mergers, neutron stars, black holes | $10^{-3}$ |

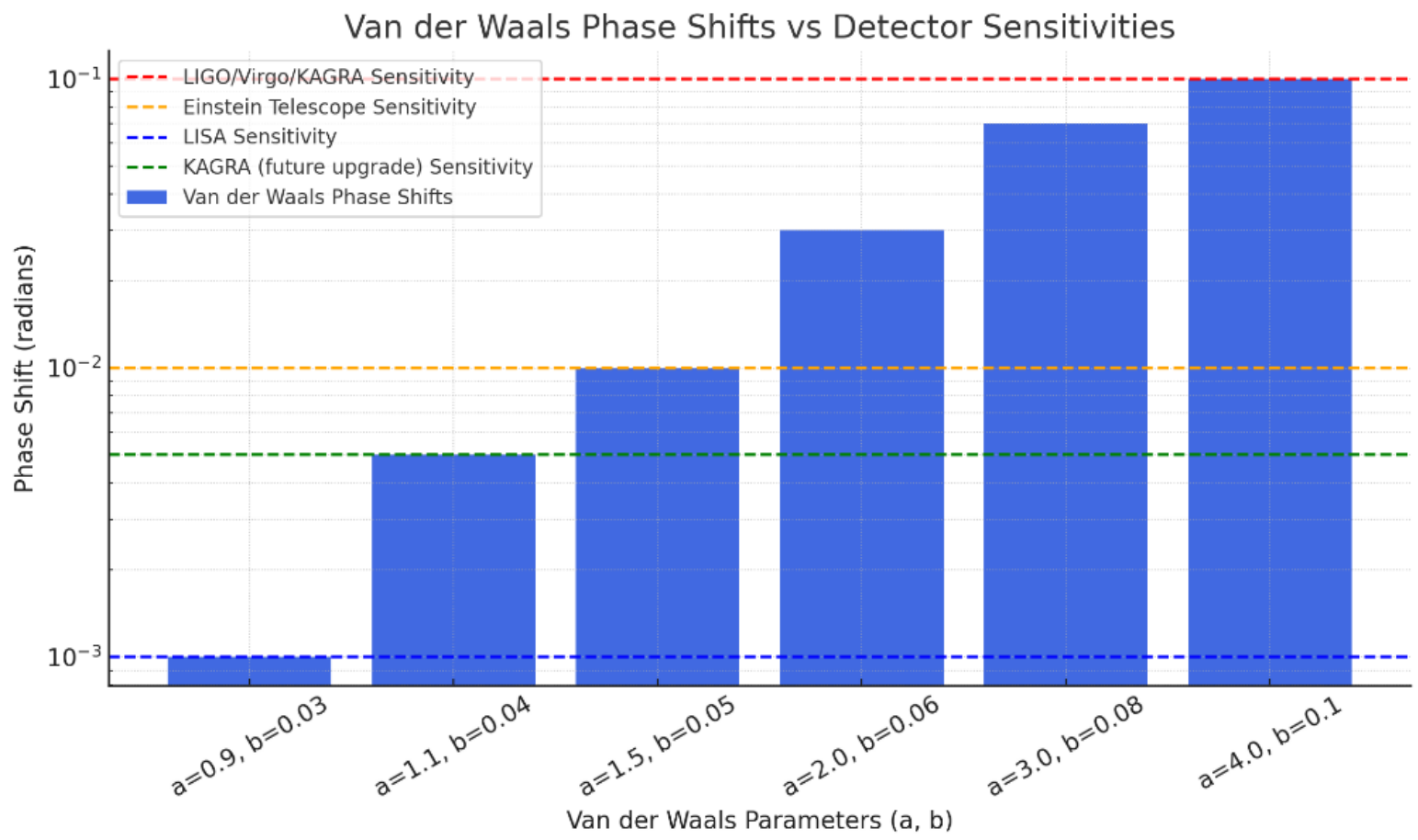

**Figure 5.** A comparison of the maximum MvdW BZT phase shift frequency dependence to the current and estimated sensitivities for the LIGO, VIRGO, KAGRA, Einstein Telescope, and LISA facilities. Details of the various EOS models are given in Appendix A.

## 4. Conclusions

We have used the nonconvex MvdW EoS to investigate the impact of the nonconvex region on the phase shift of a gravitational wave event and a GRB luminosity model to search for a detectable BZT EoS signal. In typical neutron matter, classical compressive shock waves tend to dissipate energy by heating the fluid; such action might disrupt and prevent a full QCD phase transition unless the conditions are extreme. When the fundamental derivative is negative, the compression waves will not steepen into localized shocks, but rather the waves tend to disperse and rarify upon compression producing QCD cavitation, this nonlinear anomalous shock propagation alters the phase transformation and can produce unstable metastable states very sensitive to the thermodynamic state, especially near any critical points. This behavior creates unique local variations in pressure and density which alter the conditions needed for a phase transition. This results in a coexistence region with pronounced clumping with quark clusters, nuggets, and diquark states, which can be metastable, and alternate with volumes of quark gluon plasma states with deconfined quarks. This behavior leads to potentially unusual and spatially complex phase behaviors not observed in fluid with a positive fundamental derivative. However, in a BZT neutron fluid the nonclassical shock dynamics will mean that localized energy and density fluctuations could sustain a phase transition at lower average energy levels than in standard neutron matter. This could result in a relatively low energy path to quark matter formation compared to the direct collapse model assumed for many cases. In our study, we have used the TOV equations for a spherical mass with no magnetic field, and there is no rotation; these are modifications we are currently exploring. In addition, the MvdW EoS has a natural limit to a polytropic star which provides a canonical system to investigate in parallel with the MvdW EoS. There is also extensive work on introducing the Maxwell construction to remove the nonconvex region while having the speed of sound vanish, due to no pressure gradients, while others favor the Gibbs construction to maintain chemical potentials. Both methods change the dynamics of the coexistence phase. In a similar way, one can also use an infrared regulator in the interaction term to reduce or remove the nonconvex region. We are also implementing a density-dependent interparticle force to explore a change in the van der Waals luminosity function to better match a typical GRB curve. The BZT rarefaction shocks will also modify the thermal conductivity and heat transport, thereby affecting the star's cooling behavior. The shocks will impact neutrino production in the direct Urca process: $n \rightarrow p + e^- + \bar{\nu}_e \quad p + e^- \rightarrow n + \nu_e$ and the modified Urca process: $n + n \rightarrow p + n + e^- + \bar{\nu}_e \quad p + n + e^- \rightarrow n + n + \nu_e$ where the edge density gradients should enhance neutrino production and alter neutrino emissivity, impacting potential kaon condensates. Rarefaction shocks would also reduce thermal insulation and alter the heat transport to the surface, causing a higher cooling rate and a lower surface temperature, perhaps similar to what is observed at Cas A [118].

Here we have used the MvdW EoS to examine the fundamental derivative indicating the onset of nonconvex behavior and looked at how this would induce a shift in the phase of a gravity wave-producing event. The multicomponent van der Waals (MvdW) equation of state (EOS) offers a versatile phenomenological framework for modeling dense quark matter, particularly in the context of nonconvex phase transitions relevant to neutron star interiors and gamma-ray burst (GRB) engines. However, its utility is tempered by several theoretical and practical limitations. Most notably, the MvdW EOS is not derived from first-principles QCD and lacks a direct connection to confinement dynamics or chiral symmetry breaking, which are central features of the QCD phase diagram. The model encapsulates inter-quark interactions using simplified scalar terms—namely, an attractive interaction coefficient $a_{ij}$ and an excluded volume $b_j$—which do not account for the full color, spin, or momentum-dependent structure of QCD interactions. Furthermore, thermodynamic inconsistencies can arise for certain parameter choices, including violations of causality due to superluminal sound speeds or regions of negative compressibility, unless constraints are carefully imposed. The model is typically formulated at zero temperature, omitting thermal corrections that are critical in proto-neutron stars and post-merger environments, and does not incorporate the effects of strong magnetic fields or rotation, both of which are known to significantly impact the structure and stability of compact stars. Moreover, the absence of color superconducting phases such as the color-flavor-locked (CFL) state limits the MvdW model's applicability at asymptotically high densities. Despite these limitations, the MvdW EOS remains a useful tool for exploring BZT-type nonconvex behavior and associated phenomena such as compound shock formation, especially when coupled with relativistic stellar structure calculations and gravitational wave observables.

We then examined if this could produce a measurable signature indicating the presence of internal BZT structure. We found that the formation of unusual nonclassical shock waves in a BZT hadron fluid with a negative fundamental derivative could contribute to a QCD phase transition by creating localized, high-density regions, conducive to quark gluon plasma formation along with adjacent BZT low-density rarefaction regions providing a possible means for a sustained novel phase transition. This unique behavior could facilitate the formation of quark nuggets or quark clusters, which, under the right conditions, could lead to a broader phase transition, potentially transforming the stellar core into a quark star without the need for ever higher pressures. As such, the unusual BZT properties of a hadron fluid could provide a unique mechanism by which extreme astrophysical conditions might initiate and sustain a phase change to quark matter. The formation of such an object could leave a signature in the gravity wave phase shift that could best be observed in a future gravity wave detector such as LISA or the Einstein Telescope. We are continuing to improve on this model by adding magnetic field effects, anisotropic transverse and radial pressures, coupling with an anomalous magnetic moment, and exploring the impact of nonspherical shapes.

**Author Contributions:** Formal analysis: K.A. (Keith Andrew), E.S., and K.A.A (Kristopher A. Andrew).; methodology: K.A. (Keith Andrew), E.S., and K.A.A. (Kristopher A. Andrew); investigation:

K.A. (Keith Andrew), E.S., and K.A.A (Kristopher A. Andrew).; formal analysis, numerical solutions, and visualization: K.A. (Keith Andrew), E.S., and K.A.A. (Kristopher A. Andrew); writing original draft: K.A.; writing review and editing: K.A (Keith Andrew)., E.S., and K.A.A (Kristopher A. Andrew). All authors have read and agreed to the published version of the manuscript.

**Funding:** This research received no external funding.

**Institutional Review Board Statement:** Not applicable.

**Informed Consent Statement:** Not applicable.

**Data Availability Statement:** The original contributions presented in this study are included in the article. Further inquiries can be directed to the corresponding author(s).

**Acknowledgments:** The authors wish to thank Western Kentucky University and Schlarman Academy for their kind support throughout the work on this project. All plots and numerical solutions were generated in Mathematica. The authors also wish to thank the anonymous referees who added significantly to improving this manuscript.

**Conflicts of Interest:** The authors declare no conflicts of interest.

## Appendix A. Models for Gamma-Ray Curves

To investigate the relative magnitude of the MvdW BZT noncovex region, we compare the BZT fundamental derivative values to similar modified EoS used to describe high-density phase changes. We wish to connect the microphysical properties of the interacting particles of the EOS to the observational data-based curves from astronomers. In many cases a convex EoS can be rendered nonconvex with the addition of an interaction term similar to the $a_{ij}$ interaction in the MvdW EoS. However, both the phase-changing lattice model and the symmetry breaking chiral model experience negative fundamental derivatives during their respective phase changes, and we report them as they arise in special applications similar to the one examined here. The Hadron Gas Model and the NJL model both introduce phase changes that result in a negative fundamental derivative for the quark hadron phase change and for the color-flavor-locked superconducting phase change, and we use those values for comparison. For the MIT bag model, we introduce a quark interaction term of the form $a_{ij}n_in_j$ like the MvdW EoS, which can be positive or negative depending upon the quark color factor. In this sense, we can compare the magnitude of the nonconvex regions to the MvdW case. Here we give the EOS, energy density, BZT energy and initial luminosity for a shock time scale of $\tau = 0.5$ s and a burst volume V = $10^{18}$ cm$^3$ for numerical parameter values with some variation we selected points near the midpoint and mass values are from the Particle Data Group [119], where the MvdW model uses quark constituent masses but the NJL model uses quark dynamical masses and the conditions for beta equilibrium are used in assigning values to the chemical potentials.

The MvdW system is given by [64,120]:

$$P = \sum_j \frac{n_j T}{1 - b_j n_j} - \sum_{i,j} a_{ij} n_i n_j \qquad m_u = 0.225GeV,\ m_d = 0.225GeV,\ m_s = 0.500GeV$$
$$\varepsilon = \sum_j \left( \frac{3}{2} \frac{n_j T}{1 - b_j n_j} + m_j n_j \right) + \sum_{i,j} a_{ij} n_i n_j \qquad a_{mix} = 0.4GeV^{-2} fm^3,\ b_{mix} = 0.25 fm^3,\ n_i = 2.5 n_0 \tag{A1}$$

$$MvdW : \Delta\varepsilon_{BZT} = 200 MeV fm^{-3} \quad L_0 = 6.4x10^{53} erg / s \quad .$$

The MvdW model has a wide BZT zone that supports shock-driven bursts, it produces high $L_0$, has a broad, slowly decaying GRB luminosity curve, requires careful adjustment to avoid causality violations, EOS may be too soft without constraints and it is a strong match to broad, high-luminosity GRBs like GRB 090618 and GRB 061007.

The MIT bag model with BZT level interactions is given by [121–123]:

$$P=\sum_f \frac{\mu_f^4}{4\pi^2}-B-\sum_{i,j} a_{ij} n_i n_j \qquad \varepsilon=\sum_f \frac{3\mu_f^4}{4\pi^2}+B+\sum_{i,j} a_{ij} n_i n_j \qquad \begin{aligned}&\text{Beta Equilibrium}\\ &\mu_u=325MeV\\ &\mu_d=\mu_s=450MeV\\ &B=250MeV/fm^3\end{aligned} \tag{A2}$$

$MIT:\Delta\varepsilon_{BZT}=150MeVfm^{-3} \quad L_0=4.8x10^{53}erg/s$ .

where the MIT bag Model with interactions has a harp phase transition and can produces strong Δε, has a simple analytic form, can produce a moderate-to-high $L_0$, can only support a BZT region if interactions are present, the luminosity curve is symmetric and less realistic without extension, it matches mid-width GRBs like GRB 080319B.

The NJL model is given by [124,125]:

$$P=-\left[\frac{(M-m_0)^2}{4G_S}+\frac{1}{\pi^2}\int_0^{\beta} p^2dp\left(E_p-\mu\theta\left(\mu-E_p\right)\right)\right] \qquad \varepsilon=\frac{(M-m_0)^2}{4G_S}+\frac{3}{\pi^2}\int_0^{p_F} p^2dpE_p \qquad \begin{aligned}&G_S=5.5GeV^{-2}\\ &\Lambda_{cutoff}=631MeV\\ &Bare: m_{0u,d}=5.5MeV, m_{0s}=135MeV\\ &Dynamical: M_u=M_d=335MeV, M_s=540MeV\\ &Fermi: p_{F_f=\sqrt{\mu_f^2-M_f^2}}, \quad \text{for } \mu_f>M_f\end{aligned} \tag{A3}$$

$NJL:\Delta\varepsilon_{BZT}=130MeVfm^{-3} \quad L_0=4.2x10^{53}erg/s$ .

where the NJL model has a Δε driven by chiral symmetry restoration, naturally supports color superconducting extensions, produces long tails and plateaus in the light curve, has an $L_0$ competitive with the MvdW model, matches GRBs with dual-phase decay (GRB 090618, GRB 080319B) see Appendix C for references.

The Hadron Gas Model (HGM) [126,127] in the low-temperature Boltzmann form, using modified Bessel functions of the first and second kind $K_1$, $K_2$, and restricted to the hadrons listed, where the thermodynamic contributions of each hadronic species are determined by a combination of statistical and thermal factors. The degeneracy factor $g_i$ represents the number of internal quantum states available to a species, typically incorporating spin and isospin multiplicities, $g_i = (2s_i + 1)(2I_i + 1)$ where $s_i$ is the spin and $I_i$ is the isospin of the hadron. This term scales the influence of each particle on the total pressure and energy density, with higher-spin or multi-state particles contributing more significantly. The modified Bessel functions $K_2$ and $K_1$ arise from the integration over momentum space in the partition function and play a central role in encoding the thermal behavior of massive particles. The function $K_2$ dominates the pressure and energy contributions by capturing the Boltzmann suppression of heavy species, scaling roughly as $e^{(-mi/T)}$ for $T_{mi}$ >> T. Meanwhile, $K_1$, which appears in the expression for energy density, provides a relativistic correction that accounts for the full energy content of the particles, including both kinetic and rest mass components. Together, these terms ensure that the HRG model accurately reflects the statistical mechanics of a multicomponent hadronic system in thermal equilibrium and captures the transition from relativistic to non-relativistic behavior across species and temperature ranges. The HGM is given by:

$$P=\sum_i \frac{g_i T^2 m_i^2}{2\pi^2} K_2\left(\frac{m_i}{T}\right) \qquad \varepsilon=\sum_i \frac{g_i T^2 m_i^2}{2\pi^2}\left[3K_2\left(\frac{m_i}{T}\right)+\frac{m_i}{T}K_1\left(\frac{m_i}{T}\right)\right] \qquad \begin{aligned}&\pi^0: m_{\pi^0}=0.135GeV, g=1\\ &\pi^\pm: m_{\pi^\pm}=0.140GeV, g=2\\ &K^\pm: m_{K^\pm}=0.494GeV, g=2\\ &p,n: m_{p,n}=0.938GeV, g=2\\ &\Lambda: m_\Lambda=1.116GeV, g=2\\ &\Delta: m_\Delta=1.232GeV, g=4\end{aligned} \tag{A4}$$

$HGM:\Delta\varepsilon_{BZT}=85MeVfm^{-3} \quad L_0=2.7x10^{53}erg/s$ .

where the HGM matches low-temperature lattice EOS, tends to favor a transition that produces short, sharp GRBs, in general Δε is small therefore $L_0$ is limited, so it fails to model long-duration or broad-peak GRBs, it is best for short, faint GRBs (GRB 091127); does not naturally explain long GRB luminosity tails.

The QCD lattice model is given by [128–130]:

$$
\begin{aligned}
P &= \sum_{n=0}^{N} a_{2n} T^{4+2n}, & a_0 &= 0.9, \\
& & a_2 &= 0.15 MeV^{-2}, \\
\varepsilon &= \sum_{n=0}^{N} a_{2n} \left(3+2n\right) T^{4+2n} & a_4 &= 0.02 MeV^{-4}
\end{aligned}
\tag{A5}
$$

$Lattice: \Delta\varepsilon_{BZT} = 70 MeVfm^{-3} \quad L_0 = 2.2x10^{53} erg/s$ .

where the lattice model is based directly on a QCD first-principles EOS, stability and reproducibility are good for calibration and EOS matching, produces a small Δε and soft peak luminosity, no BZT region unless effective model interactions are added, can approximate GRB plateaus but underpredicts $L_0$ for bright GRB events, the lattice model is most effective for events like GRB 111228.

Each dense matter model offers unique strengths and limitations in modeling quark star interiors and their connection to BZT shock dynamics and GRB luminosity profiles. The multicomponent van der Waals (MvdW) EOS provides a tunable framework that naturally supports nonconvex thermodynamic behavior through its attractive and repulsive interaction parameters, making it highly effective for modeling BZT-type shocks and generating strong, broad GRB luminosity curves. However, it lacks a direct derivation from QCD and may violate causality if not properly constrained. The MIT bag model, while analytically simple and widely used in hybrid star constructions, inherently lacks interaction terms and BZT structure unless explicitly modified; its GRB output tends to be more symmetric and shorter in duration. The Nambu–Jona-Lasinio (NJL) model captures essential QCD features such as chiral symmetry breaking and restoration, and it can produce BZT-like effects near the chiral critical point, giving rise to dual-phase or delayed GRB light curves. Its main weakness lies in the absence of confinement and its sensitivity to regularization. The Hadron Resonance Gas (HRG) model accurately reproduces low-density hadronic thermodynamics and matches lattice results below the crossover, but its limited ability to model phase transitions and lack of nonconvexity restrict its use in BZT shock applications, leading to fast-fading, less energetic GRB profiles. Finally, lattice QCD provides the most fundamental insight into the QCD equation of state at finite temperature and low baryon density, yet the sign problem hampers its application at neutron star core densities. It predicts smooth crossovers without strong shocks, producing brief and relatively featureless GRB signals unless supplemented with effective models. In sum, the ability of each EOS to generate observable BZT shocks and realistic GRB luminosity curves is directly tied to its capacity to support nonconvex thermodynamics and a robust first-order phase transition.

To model the gamma-ray burst (GRB) light curves associated with phase transitions in quark stars, we introduce a semi-analytic energy release framework based on the propagation of a shock front through a nonconvex Bethe–Zel'dovich–Thompson (BZT) region using the Rees Double Power Law Luminosity Function for GRB (also known as the broken power law) phenomenological model [131–133] with the Li Shock Breakout GRB Model for the BZT shockwave [134–136]. As the shock traverses the BZT layer, the local drop in the fundamental derivative G leads to a release of internal energy, which is partially converted into observable electromagnetic emission and structure changes that alter gravity wave production. The luminosity for each model is described using a standard dual gamma-distribution-shaped function of time:

$$L(t)=\begin{cases} L_0\left(\dfrac{t}{t_0}\right)^{-a_1} & , \quad t<t_b \\ L_0\left(\dfrac{t_b}{t_0}\right)^{-a_1}\left(\dfrac{t}{t_b}\right)^{-a_2} & , \quad t\geq t_b \end{cases} \tag{A6}$$

where $L_0$ is the peak luminosity, $t_b$ is the characteristic rise time, and n is the shape parameter controlling the rapidity of onset. The two exponents $a_1$ and $a_2$ are used to match the observed GRB, as shown in Table A1 below. The associated measured $L_0$ values are matched to the relative strength and spatial extent of the BZT region for each EOS model. To model gamma-ray burst (GRB) light curves from first principles, we apply the Rees broken power law luminosity function, parameterized as $L_0$ is the peak luminosity, $t_0$ is the initial normalization time, $a_1$ and $a_2$ are the early and late decay indices, and $t_b$ is the blast or break time marking the transition between emission regimes. We constrain $L_0$ using the microphysical energy density discontinuity $\Delta\epsilon$ predicted by various equations of state (EOS), including $L_0 \approx \Delta\epsilon \cdot V/\tau$, with V the emitting volume and $\tau$ the release timescale. For example, using EOS-derived values of $\Delta\epsilon \sim 2 \times 10^{35}$ erg/cm$^3$ and $\tau = 0.5$ s, the MvdW model yields $L_0 \sim 4 \times 10^{53}$ erg/s, consistent with Swift-XRT luminosity data for GRB 090618 (URL accessed 5/12/2024: https://www.swift.ac.uk/xrt_curves/ ). We fit $t_b$ and decay slopes $a_1$, $a_2$ to the observed light curve, obtaining best-fit values $t_b \sim 4 \times 10^4$ s, $a_1 \sim 0.8$, and $a_2 \sim 1.7$, which match the observed transition from engine-driven plateau to afterglow phase. This approach bridges microphysical EOS modeling and GRB data, enabling direct EOS constraints from light curve morphology. For example, the multicomponent van der Waals (MvdW) model, which exhibits the broadest and most intense BZT zone, yields the largest $L_0$ and a slower decay (larger $t_0$), while the Hadron Gas Model—featuring only a narrow BZT window near saturation density—produces a sharp, lower-luminosity pulse. By selecting $L_0$ in proportion to the integrated energy release, the Li Shock Model as applied here can be expressed as:

$$L_0=\frac{E_{shock}}{\tau}=\frac{\int_{V_{BZT}}\Delta\varepsilon\, dV}{\tau}\approx\frac{\Delta\varepsilon_{BZT}V}{\tau} \tag{A7}$$
$$E=\int\varepsilon\, dV, \quad 10^{-3}s<\tau<10^{3}s, \quad V\approx 10^{18}cm^{3}$$

where $\Delta\epsilon_{BZT} \sim \alpha|\Delta G|$ and scaling the time response to the shock transit time, this formulation provides a unified way to represent GRB signatures from different QCD phase transition scenarios described by an EOS, linking fluid nonconvexity to observable high-energy astrophysical transients.

We use the data from Ruffini and Izzo [137] for the smoothed and idealized GRB curves to match the model with parameters given in Table A1.

**Table A1.** GRB data and model fit from Ruffino and Izzo [137] to compare to the luminosity curves for the MvdW EOS.

| | Detector | Frequency Range (Hz) | Sensitivity (Strain) | SNR Threshold | Application | Phase Shift Sensitivity |
|---|---|---|---|---|---|---|
| 1 | LIGO | 10-1000 | $\sim 10^{-23}$ | ~8 | Binary mergers, neutron stars, black holes | $10^{-2}$ |
| 2 | Virgo | 10-1000 | ~$10^{-23}$ | ~8 | Binary mergers, neutron stars, black holes | $10^{-2}$ |
| 3 | LISA | 0.01-1 | ~$10^{-21}$ | ~1 | Supermassive black holes, early universe | $10^{-5}$ |
| 4 | Einstein Telescope | 1-1000 | ~$10^{-24}$ | ~5 | Binary mergers, cosmological signals | $10^{-4}$ |
| 5 | KAGRA | 10-1000 | ~$10^{-23}$ | ~8 | Binary mergers, neutron stars, black holes | $10^{-3}$ |

Given the values of $a_1$ and $a_2$ from observational data the EOS is used to match the observed $L_0$ using $\Delta\varepsilon$ and then plotted using the Rees Double Power Law Luminosity Function for GRBs, these are shown in Figure 4. these end up being smoothed curves with a clear phase change point that does not exhibit the observational variation and scatter associated with the real data. In Figure A1, we show a plot of some of the points for GRB 090618 with a Rees Double Power Law fit and the corresponding MvdW parameter fit in the $a_{mix}$–$b_{mix}$ region.

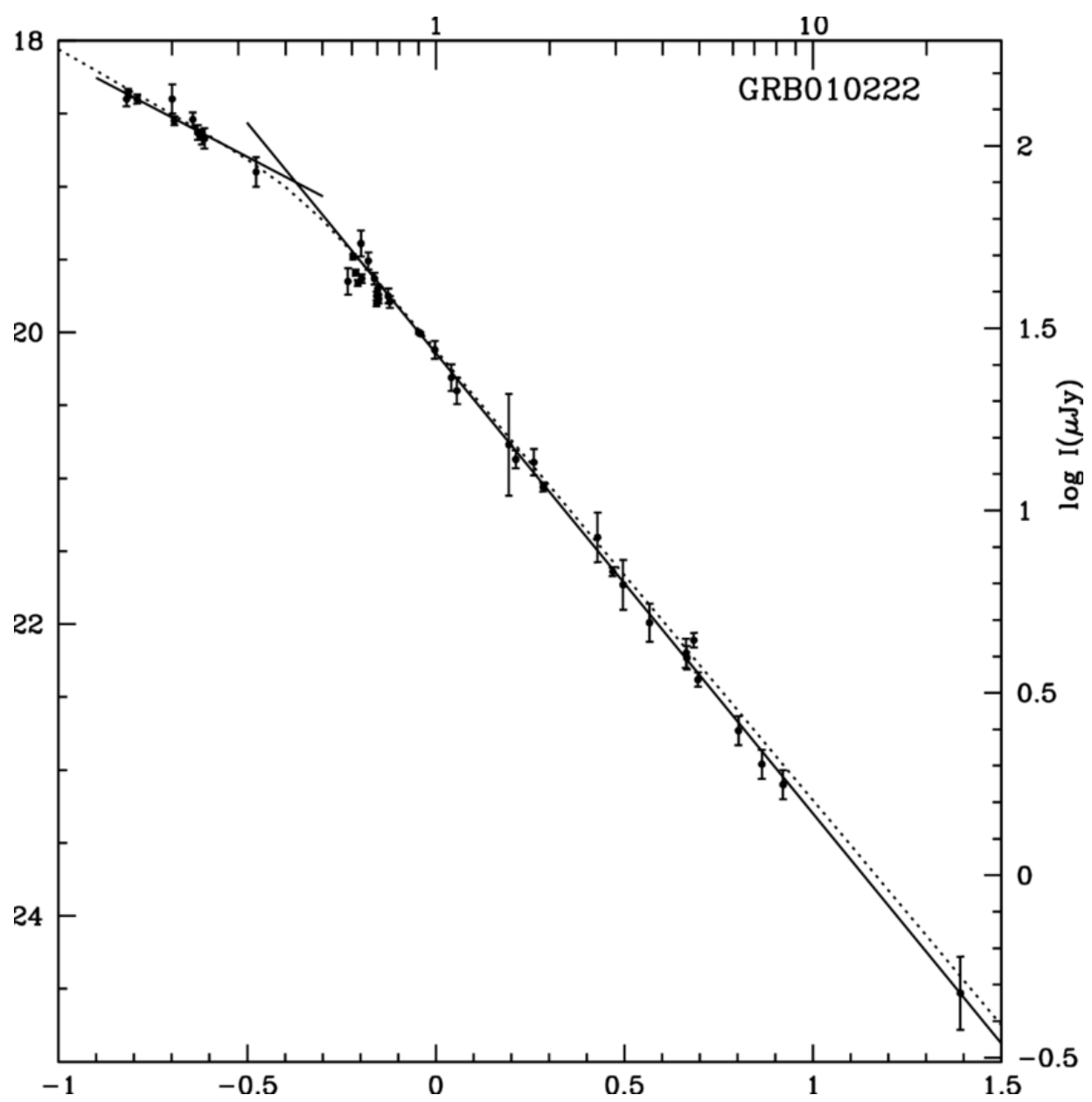

**Figure A1.** The GRB luminosity curve (subset of points) fit to the Rees broken power law model light curve of GRB 010222 from Cowsik, et al. using the Indian Astronomical Observatory, Hanle, and the telescopes at the Vainu Bappu Observatory, Kavalur, with a broken power law fit to the data, where the data, figure and broken power law fit are all from the Cowsik group [138,139]. The underlying MvdW model then estimates a peak luminosity $L_0 = 4.2 \times 10^{45}$ erg/s, a reference time $t_0 = 100$ s, early-time decay index $a_1 = 0.542$, late-time decay index $a_2 = 1.263$, and a blast break time at $t_b = 3.71 \times 10^4$ s. Observational points are indicated by the markers. This GRB light curve can accommodate a MvdW solution as shown in the next figure.

To complement the broader compactness and observational survey plot presented in Figure 2, we include here a focused viability map in Figure A2, of the ($a_{mix}$, $b_{mix}$) parameter space that highlights the intersection of theoretical and observational constraints. This diagnostic plot narrows in on the quantum-consistent region for the MvdW EOS and overlays contours of constant compactness alongside the BZT region, defined by a negative fundamental derivative. Unlike the broader color-mapped gradient in Figure 2, this visualization cleanly distinguishes the boundary of the BZT zone and clarifies how nonconvex thermodynamic behavior depends on the balance of attractive and repulsive interactions for an energy change $\Delta\epsilon$. The color map itself displays the effective energy discontinuity $\Delta\epsilon$, a quantity directly linked to both shock dynamics and observable GRB luminosity. Compactness curves are labeled and overlaid, allowing immediate comparison to observational thresholds from neutron stars and gravitational wave events. A red circle identifies a viable model for GRB 010222 using the MvdW EOS that satisfies causality, compactness, and quantum bounds simultaneously. By constraining attention to a physically viable domain and highlighting the synergy between microscopic EOS parameters and macroscopic observables, this plot serves as a precise guide to the optimal MvdW parameter space and illustrates where realistic shock-induced phase transitions may occur.

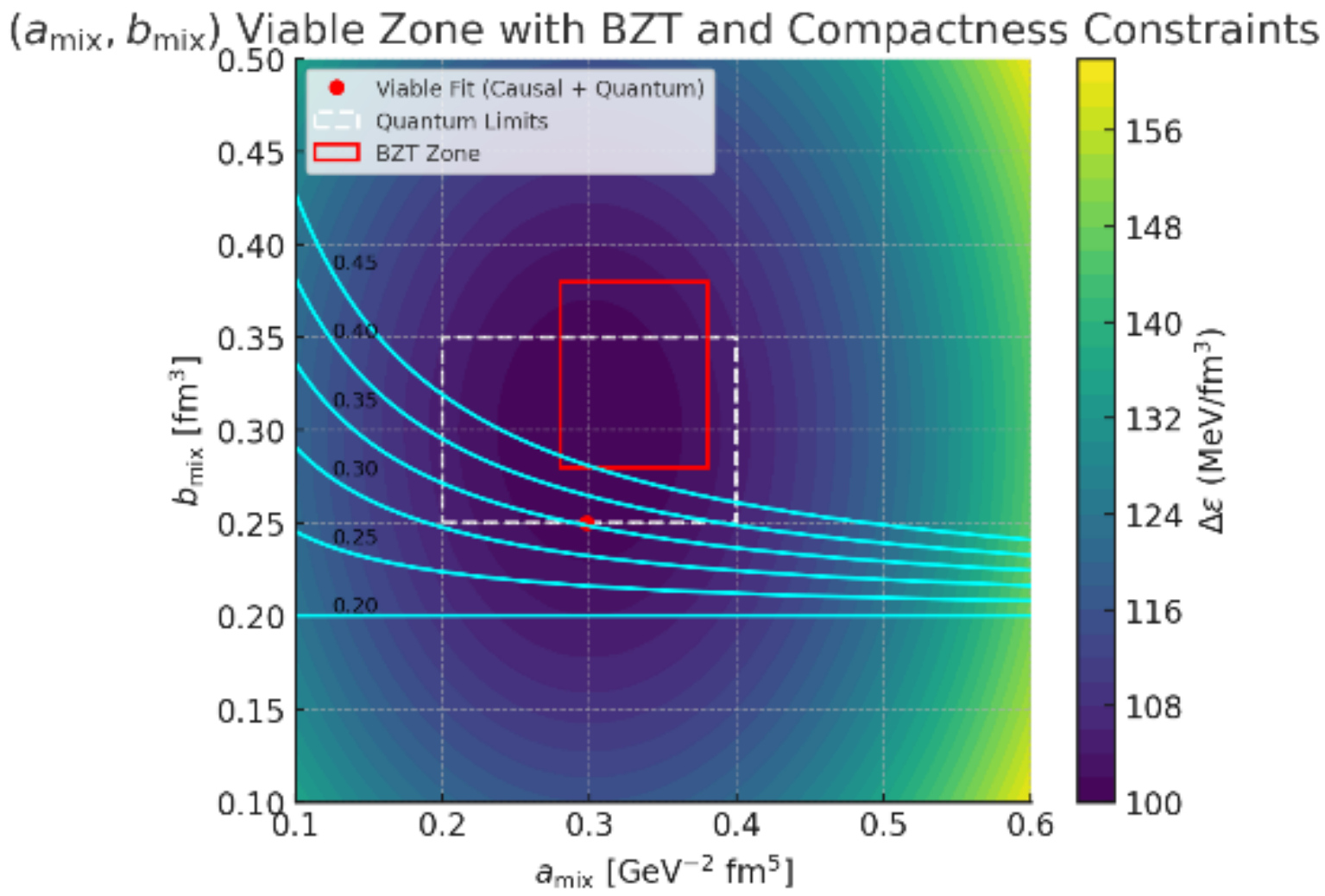

**Figure A2.** Contour plot of the effective energy shift $\Delta\epsilon$ ($a_{mix}$,$b_{mix}$) in units of MeV/fm$^3$ for the multicomponent van der Waals (MvdW) equation of state, mapped over a range of interaction parameters. The red circle marks the best-fit point for GRB 010222, derived from its observed afterglow slopes ($a_1$ = 0.542, $a_2$ = 1.263 from Cowsik) and constrained to lie within the quantum-limited region (white dashed rectangle) and below the causal compactness threshold ($C < 0.354$). The BZT zone (solid red box) indicates where nonconvex thermodynamic behavior allows for Bethe–Zel'dovich–Thompson-type shock propagation. Compactness contours from C = 0.20 to 0.450 are shown in cyan, with values labeled in black on the left. The selected point has a compactness of approximately $C \approx 0.34$ and a local energy shift of $\Delta\epsilon \approx 101.0$ MeV/fm$^3$, positioning it just outside the BZT region but within the physically viable range for dense quark matter sources consistent with GRB observations.

## Appendix B. Bayesian Inference Corner Plot for MvdW

This corner plot visualizes the posterior distributions of key MvdW EOS parameters—namely, the attractive interaction strength $a_{mix}$, excluded volume $b_{mix}$, and a derived phase-sensitive quantity $\Delta\phi = |\Delta\Psi|$. The 1D histograms reveal the most probable parameter values, while the 2D contours illustrate correlations and trade-offs, such as the inverse relationship between $a_{mix}$ and $b_{mix}$ required to maintain EOS consistency with astrophysical constraints. This analysis provides insight into how microphysical quark interactions influence macroscopic observables, such as tidal deformability or BZT-driven GRB signals, and helps identify the viable region of parameter space for modeling phase transitions in quark stars.

This figure presents the posterior distributions and parameter correlations for a multicomponent van der Waals (MvdW) equation of state (EOS) used in modeling dense matter, likely within the context of neutron stars or quark stars. It is a corner plot showing marginalized 1D histograms along the diagonal and 2D posterior contours off-diagonal for three key EOS parameters: $a_{mix}$: the effective attractive interaction strength between quark flavors or particle species in the MvdW framework; bmix: the effective excluded volume per particle, encoding short-range repulsion; and $|\Delta\Psi|$: the gravity wave phase shift resulting in an observable signature.

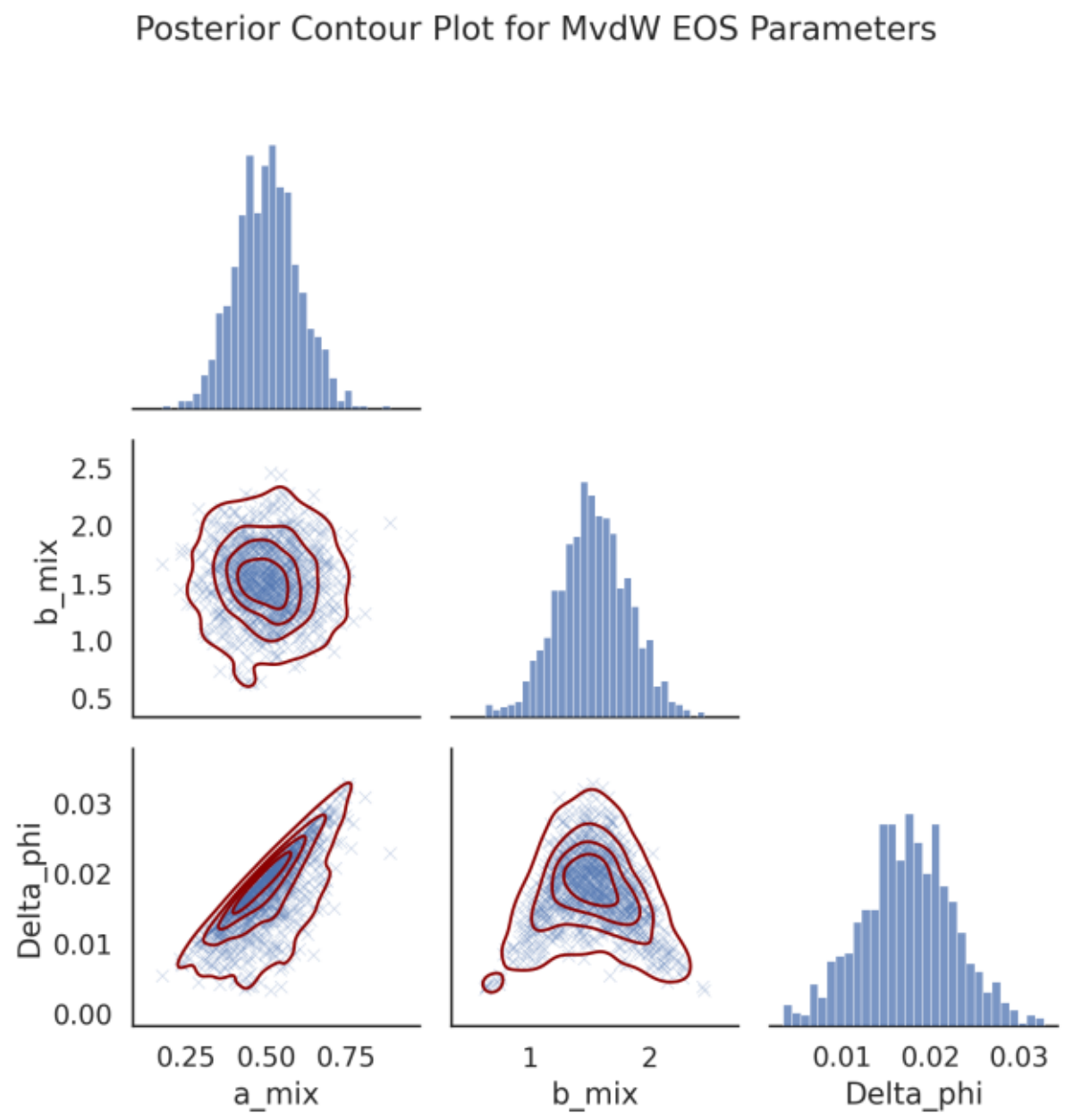

**Figure A3.** Posterior contour and marginal distribution plot for multicomponent van der Waals EOS parameters derived from Bayesian inference. Shown are the marginalized distributions for the attraction parameter $a_{mix}$, excluded volume parameter $b_{mix}$, and derived observable shift $|\Delta\Psi|$. The central red contours reflect 68%, 95%, and 99% confidence intervals, revealing correlations among EOS parameters relevant for compact star structure or gravitational wave phase evolution.

The 2D contours represent confidence regions (set to 68%, 95%, and 99%) showing the most probable combinations of parameters inferred from data (e.g., mass–radius measurements, tidal deformabilities, or gravitational wave phase shifts). There is a strong positive correlation between $a_{mix}$ and $|\Delta\Psi|$, and a moderate anticorrelation between $b_{mix}$ and $|\Delta\Psi|$, indicating how changes in interaction strength or excluded volume affect observable features.

## Appendix C. Neutron Star Sources

In addition to the Swift Database, the neutron stars used in Figures 2 and 3 are from the following references:

**Table A2.** Additional references for the neutron stars used in Figures 2, 3, and A2 and model calculations in Appendix A.

| Neutron Star | Reference |
|---|---|
| PSR J0030+0451 | [140] |
| PSR J0740+6620 | [141] |
| PSR J1614–2230 | [142] |
| PSR J0952–0607 | [143] |
| PSR J0348+0432 | [144] |
| PSR J2124–3358 | [145] |
| PSR J1909–3744 | [146] |
| GRB170817A | [147] |
| GRB190425 | [148] |
| GRB211211A | [149] |
| GRB230307A | [150] |
| GRB221009A | [151] |
| GRB150101B | [152] |